\documentclass[9pt,twocolumn,twoside]{opticajnl}

\journal{optica} 

\setboolean{shortarticle}{false}

\hypersetup{draft}

\usepackage{lineno}

\title{High-speed mid-infrared single-photon upconversion spectrometer}

\author[1]{Tingting Zheng}
\author[1,3,5,*]{Kun Huang}
\author[1]{Ben Sun}
\author[1]{Jianan Fang}
\author[2]{Yongyuan Chu}
\author[2,$\dagger$]{Hairun Guo}
\author[1,3]{E Wu}
\author[1,3]{Ming Yan}
\author[1,3,4,6,7]{Heping Zeng}

\affil[1]{State Key Laboratory of Precision Spectroscopy, East China Normal University, Shanghai 200062, China}
\affil[2]{Key Laboratory of Specialty Fiber Optics and Optical Access Networks, Shanghai University, Shanghai 200444, China}
\affil[3]{Chongqing Key Laboratory of Precision Optics, Chongqing Institute of East China Normal University, Chongqing 401121, China}
\affil[4]{Chongqing Institute for Brain and Intelligence, Guangyang Bay Laboratory, Chongqing 400064, China}
\affil[5]{Collaborative Innovation Center of Extreme Optics, Shanxi University, Taiyuan, Shanxi 030006, China}
\affil[6]{Jinan Institute of Quantum Technology, Jinan, Shandong 250101, China}
\affil[7]{Shanghai Research Center for Quantum Sciences, Shanghai 201315, China}
\affil[*]{khuang@lps.ecnu.edu.cn}
\affil[$\dagger$]{hairun.guo@shu.edu.cn}

\begin{abstract}
Sensitive and fast mid-infrared (MIR) spectroscopy is highly attractive in a variety of applications including astronomical observation, pharmaceutical synthesis, and environmental monitoring. However, the performance of conventional MIR spectrometers has long been hindered by the limited sensitivity of narrow-bandgap detectors and/or the deficient brightness of broadband light sources. Here, we devise and implement an ultra-sensitive and broadband MIR upconversion spectrometer, which integrates a supercontinuum source covering 1.5-4.2 $\mu$m based on a silicon nitride nanophotonic waveguide. High-efficiency and low-noise nonlinear frequency upconversion is realized based on coincidence pulsed pumping with spectro-temporal optimization, which enables to leverage silicon detectors for facilitating MIR single-photon spectroscopy at 0.2 photons/nm/pulse. Furthermore, the upconversion-based array spectrometer is manifested with high-speed spectral acquisition rates beyond 200 kHz, which is about ten-fold faster than the state-of-the-art scan rates for FTIR-based spectrometers at a comparable spectral resolution. The achieved features of broadband spectral coverage, single-photon sensitivity, and sub-MHz refreshing rate might open up new possibilities in infrared transient spectral measurements in combustion analysis, high-throughput sorting and reaction tracking, among others. 
\end{abstract}

\setboolean{displaycopyright}{false}

\begin{document}

\maketitle

\section{Introduction}
Mid-infrared (MIR) spectroscopy in the spectral range from 2.5 to about 25 $\mu$m provides a non-destructive and label-free analytical technique by addressing the molecular rotational-vibrational transitions, which has widely been used for qualitative and quantitative chemical characterization of a variety of substances in industrial, environmental, and biomedical applications \cite{EbrahimZadeh2008Book, Vodopyanov2020Book, Pilling2016CSR}. Commonly, the analysis of MIR absorption spectrum is realized by the Fourier-transform infrared spectroscopy (FTIR) due to the features of broadband spectral coverage and high spectral resolution \cite{Griffiths1983Science}. However, the temporal resolution of conventional FTIR spectrometers is severely restricted by the low scan rate of mechanical delay lines \cite{Hashimoto2018NC}. The limited spectral acquisition rate imposes a major challenge to approach the high-speed MIR spectroscopy, which is pertinent to complex and dynamic scenarios such as gaseous combustion analysis, high-throughput sorting, and biochemical reaction monitoring \cite{Werblinski2017AO, Hiramatsu2019SA, Herman2019JMS}. To this end, advanced designs based on ultrasonic resonators \cite{Suss2016RSI} or phase-controlled delay lines \cite{Hashimoto2021LPR} are proposed to implement rapid-scan FTIR instruments, which can improve the temporal resolution up to tens of kHz, albeit with reduced spectral resolutions and/or narrowed spectral bandwidths \cite{Griffiths1999VS}. Notably, emerging MIR dual-comb spectroscopy allows one to acquire interferograms in a mechanical-scan-free fashion \cite{CosselJOSAB2016, Schliesser2012NP}, but the acquisition speed for obtaining a sequential of spectra with high signal-to-noise ratios (SNRs) is practically limited by the data-averaging process due to the low pulse energy, especially for broadband comb sources \cite{Schliesser2012NP, Muraviev2018NP, Ycas2018NP}. Hence, it remains a long-sought-after goal to realize high-contrast MIR spectroscopy with fast speed, broadband coverage, and high resolution.

Alternatively, dispersive infrared spectrometers based on diffractive gratings and array detectors offer the potential for high-speed measurements because all the spectral elements can be captured simultaneously over a wide spectral range \cite{Nugent2012OL, Iwakuni2019OE}. However, the available MIR focal plane arrays (FPAs) based on narrow-bandgap semiconductors like MCT, InSb, or PbSe currently suffer from several technical limitations such as high dark noises, small pixel numbers, and low frame rates \cite{Razeghi2014RPP}, which impose restrictions on the detection sensitivity, spectral resolution and measurement speed \cite{Hamm1994OL, Arrivo1997PL}. For instance, array-based MIR spectrometers have been demonstrated at a frame rate of 390 Hz based on a PbSe detector with 160 elements \cite{Ji2004RSI}. Moreover, the MIR detectors are often required to operate at cryogenic conditions, thus resulting in additional complexity and cost \cite{Wang2019Small}. In this context, MIR upconversion spectrometers have been proposed to leverage high-performance silicon pixelated sensors by spectrally converting the MIR spectral information into the visible or near-infrared bands \cite{Neely2012OL, Zhu2012OE, Lichtenberg2016JOSAB,  Rodrigo2021LPR}. 

Pioneering implementation of MIR upconversion spectrometers can be traced back to 1970s \cite{Gurski1978AO}, yet the detection efficiency is prohibitively small especially for a broad operation bandwidth \cite{Heilweil1989PL, DeCamp2005OL}. To date, tremendous efforts have been dedicated to expanding the wavelength acceptance window for efficient nonlinear frequency upconversion, for instance by resorting to vectorial phase-matching conditions \cite{Barh2017OL, Israelsen2019LSA}, chirped-poling adiabatic interactions \cite{Jahromi2019OE, Friis2019OL, Mrejen2020LPR}, multi-channel pumping sources \cite{Lichtenberg2016JOSAB}, or cascaded-stacking nonlinear crystals \cite{Rodrigo2021LPR}. Recently, the upconversion-enabled array spectrometer has demonstrated kilohertz spectral acquisition rates for a wavenumber range about 600 cm$^{-1}$ \cite{Wolf2017OE}. It is worth noting that emerging products are available to provide an impressive spectrum readout rate up to 130 kHz \cite{NLIR}, albeit that the signal-to-noise ratio for the acquired spectrum may be limited by the detection sensitivity at about 25 pW/nm. It is thus favorable to improving the high-speed and high-contrast spectral acquisition by further enhancing the detection sensitivity of the MIR upconversion spectrometer, where sufficient photons can be accumulated within a shorter snapshot time \cite{Junaid2019Optica}. However, it is still a challenging task to realize a high conversion efficiency and a low background noise for a broadband spectral coverage. To date, the single-photon operation in the framework of parametric spectral transduction has only been demonstrated for narrow spectral bandwidths of about tens of nm \cite{Dam2012NP,Huang2022NC}.

In addition to the high detection sensitivity, MIR sources with high spectral brightness are required to implement high-speed and high-SNR spectroscopic measurements \cite{Jahromi2019OE, Borondics2018Optica, Zorin2022OE}. Although single-photon MIR upconversion spectrometers have been reported by using illumination sources based on thermal radiation \cite{Dam2012NP} and parametric fluorescence \cite{Kalashnikov2016NP, Vanselow2020Optica}, yet the intrinsically low spectral radiance from black-body emission or spontaneous down conversion makes the upconversion spectrometer incompatible with the high-speed operation. Therefore, it is highly demanded to simultaneously realize the high-brightness source, broadband frequency transduction, and sensitive upconversion detection, with an aim to fully demonstrate the agile potential of MIR upconversion spectrometers.

Here, we devise and implement a broadband MIR upconversion spectrometer with single-photon detection sensitivity and sub-MHz refreshing rate. The involved high-brightness illumination is accessed with a nanophotonic supercontinuum laser source based on a silicon nitride (Si$_3$N$_4$) waveguide. Specifically, the photonic chip-based supercontinuum is directly driven by an erbium-doped fiber laser at 1.55 $\mu$m, and thus provides a turn-key, coherent, and compact MIR source with a spectral coverage from 1.5 to 4.2 $\mu$m. The prepared MIR pulses are then pumped by a passively-synchronized ytterbium-doped fiber laser at 1.03 $\mu$m within a chirped-poling lithium niobate crystal, which permits a single-shot broadband conversion without the need of parameters tuning. Moreover, the coincidence-pumping configuration contributes to increase the conversion efficiency with the high peak power, and meanwhile to suppress the background noise via the ultrashort optical gating. Consequently, a noise equivalent power about 3 fW/Hz$^{1/2}$ is achieved, which permits ultra-sensitive MIR spectroscopy at the single-photon-level illumination about 0.2 photons/nm/pulse. The combination of superior detection sensitivity and bright source radiance allows us to capture high-contrast spectra with an integration time down to microseconds. In the experiment, an unprecedented spectral acquisition rate up to 212,500 frames per second is achieved at a spectral resolution of 5 cm$^{-1}$, which is over one order of magnitude faster than reported results at comparable signal-to-noise ratios. The presented high-speed MIR spectrometer would be useful to investigate non-repetitive transient phenomena and highly dynamic processes.

\section{Experimental Setup}

\begin{figure*}[t!]
\centering
\includegraphics[width=0.65 \textwidth]{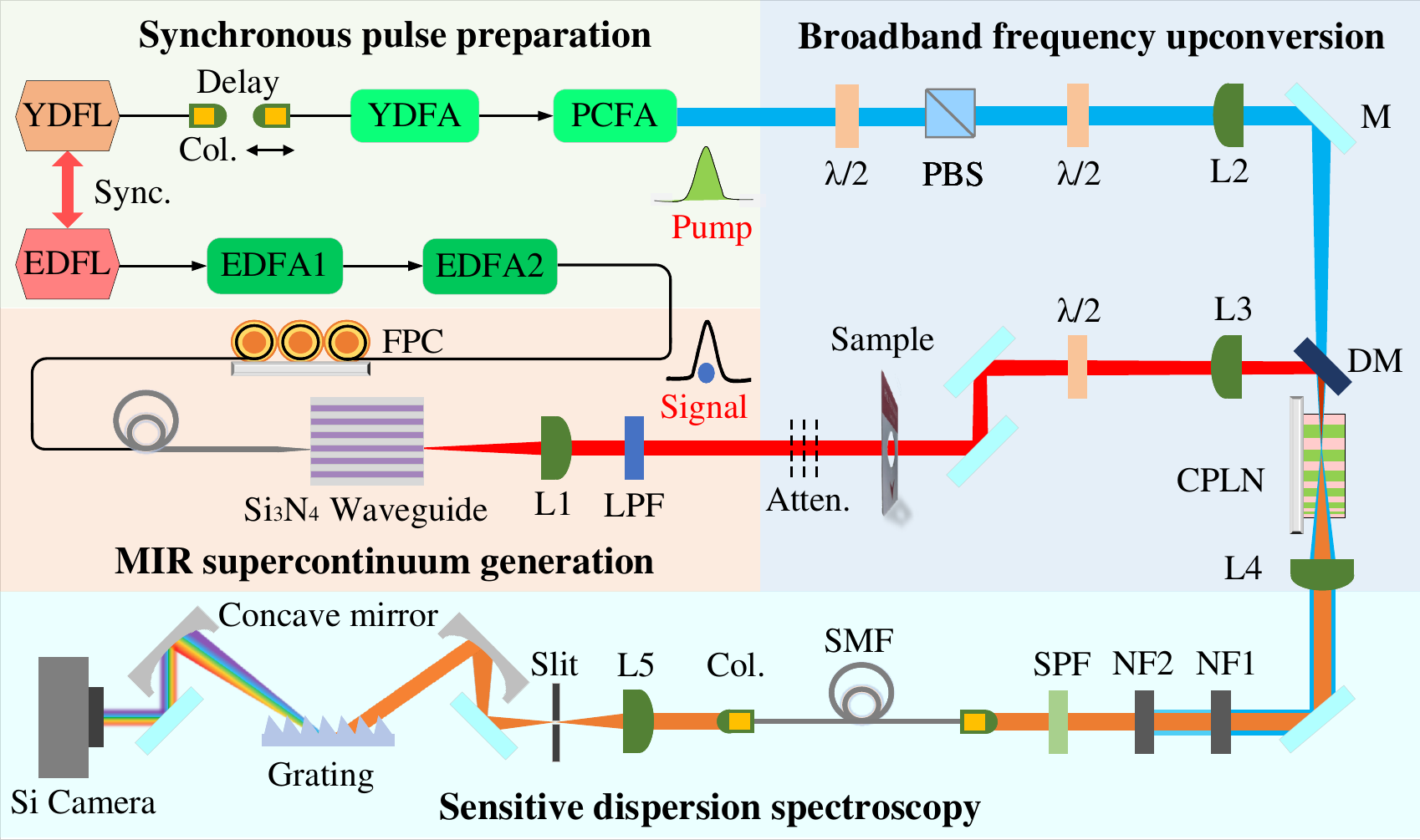}
\caption{Experimental setup for broadband and sensitive MIR upconversion spectrometer based on the coincidence-pumping configuration. The involved light sources originate from passively synchronized Er- and Yb-doped fiber lasers (EDFL and YDFL), which deliver dual-color mode-locked ultrafast pulses at an identical repetition rate. The EDFL output at 1550 nm is boosted by two cascaded fiber amplifiers (EDFAs) before being injected into a silicon nitride waveguide for direct MIR supercontinuum generation. The generated high-brightness light source is collimated into the free space to illuminate test samples. Subsequently, the transmittance is spatially combined with the synchronous narrow-band pump pulse at 1030 nm by a dichroic mirror (DM). The mixed beams are then steered into a chirped-poling lithium niobate (CPLN) crystal to facilitate broadband sum frequency generation. The upconverted light passes through a series of spectral filters before being coupled into a single-mode fiber (SMF). Finally, the filtered beam is sent into a high-performance grating spectrometer based on silicon cameras, which enables to realize fast and sensitive spectroscopy. EDFA1, EDFA2: Er-doped fiber amplifier; YDFA: Yb-doped fiber amplifier; PCFA: photonic crystal fiber amplifier; Col: fiber collimator; PBS: polarization beam splitter; FPC: fiber polarization controller;  Atten: neutral density attenuator; $\lambda$/2: half-wave plate; L: lens; M: mirror; SPF and LPF: short- and long-pass filter; NF: notch filter.}
\label{fig1}
\end{figure*}

Figure \ref{fig1} presents the experimental scheme for the high-performance MIR upconversion spectrometer, which consists of initial light source preparation, broadband frequency upconversion, and sensitive dispersive spectroscopy. The involved light source in the experiment originates from a passively synchronized fiber laser system, which is comprised of an Er-doped fiber laser (EDFL) and an Yb-doped fiber laser (YDFL) at the repetition rates of 97 MHz and 19.4 MHz, respectively. The two fiber lasers are arranged in a master-slave configuration, where the master pulses from the EDFL are injected into the optical cavity of the slave YDFL to perform the cross-phase modulation (XPM). Consequently, the two mode-locked fiber lasers are passively synchronized without stringent requirements of high-speed feedback electronics, thus providing a simple yet robust way to obtain dual-color pulses at disparate wavelengths \cite{Huang2018OE}. Note that the synchronous laser sources have also been realized based on electronical timing gating \cite{Vandevender2004JMO} or optical parametric oscillators \cite{Junaid2019Optica}. More details about the implementation for the all-optical synchronization can be found in Supplementary Note 1. In our experiment, the repetition rate of the YDFL is designated at the fractional harmonic of the EDFL by taking into account of the synchronous stability and pulse properties, with an aim to improve the performance of the subsequent frequency upconversion. Indeed, a lower repetition rate leads to a longer cavity length to enhance the XPM effect, and meanwhile the smaller duty cycle of the pulse train favors to obtain a higher peak power.

\begin{figure*}[t!]
\centering
\includegraphics[width=0.64\textwidth]{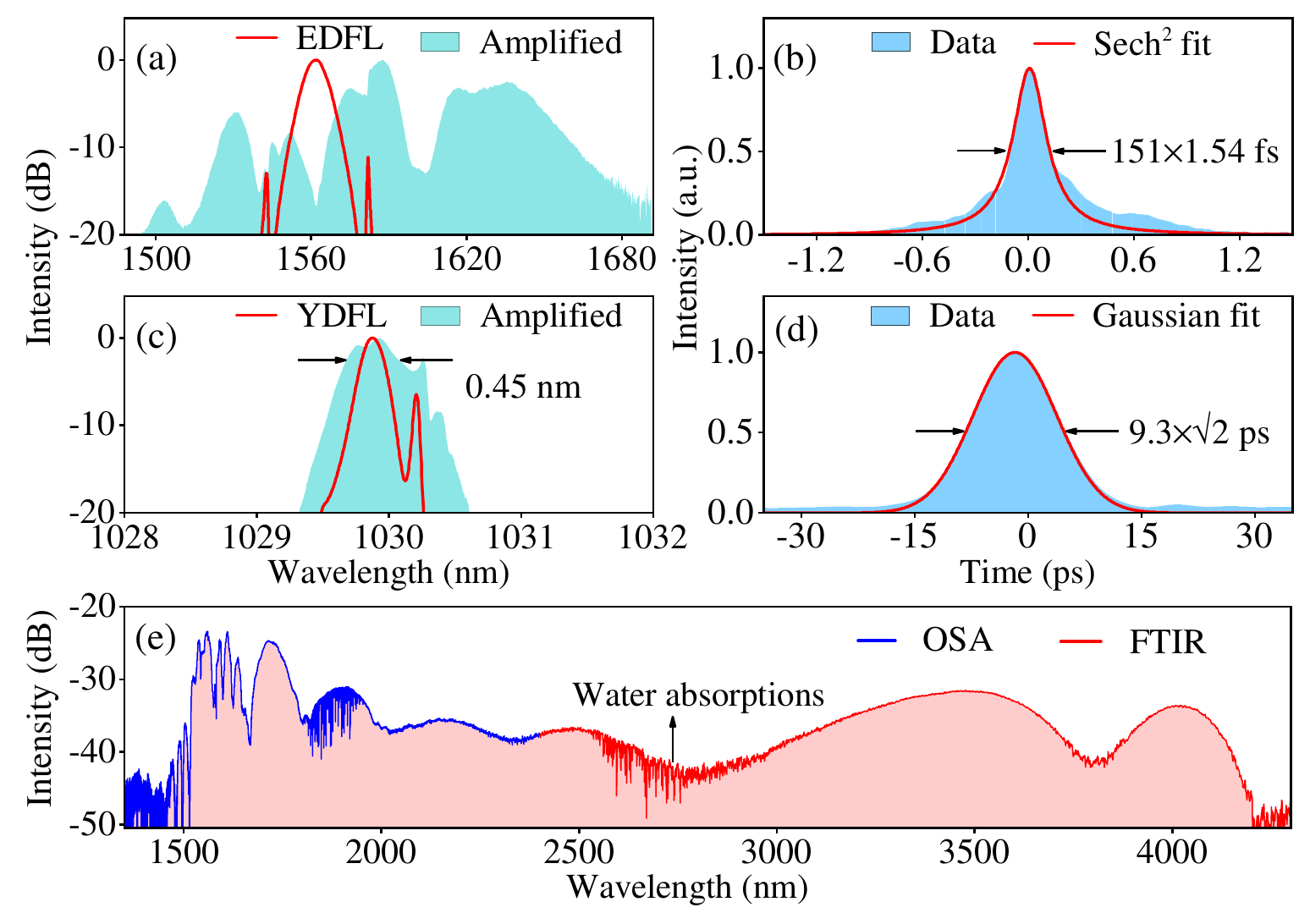}
\caption{Spectro-temporal characterization of involved optical pulses. (a) Measured spectra for the EDFL and the amplified output by an optical spectral analyzer (OSA). (b) Measured auto-correlation trace for the amplified light after the EDFAs. Scaling factor of 1.54 is taken to deduce the actual pulse duration with the assumption of a Sech$^2$ profile. (c) Optical spectra for the YDFL and the amplified output. The 3-dB bandwidth is estimated to be 0.45 nm. (d) Measured auto-correlation trace for the amplified light after the YDFAs. Scaling factor of $\sqrt{2}$ is used to infer the actual pulse duration under the Gaussian-profile assumption. (e) Experimentally observed supercontinuum in a silicon nitride waveguide, measured with a combination of an OSA (blue line) and a FTIR spectrometer (red line).}
\label{fig2}
\end{figure*}

The output of the EDFL at 1560 nm is boosted by two-stage fiber amplifiers with normal dispersion, which results in a broaden spectrum as shown in Fig. \ref{fig2}(a). Temporal compression is implemented by using a section of single-mode fiber to obtain a pulse duration of 151 fs as inferred from the measured auto-correlation trace in Fig. \ref{fig2}(b). The energetic pulse with a peak power about 10 kW is coupled into a nanophotonic Si$_3$N$4$ waveguide with the help of a fiber tip. The height and width of the waveguide section are 0.82 and 2.67 $\mu$m, which are critical to tailor the overall dispersion of the waveguide \cite{Guo2018NP, Grassani2019NC}. The synergic effects of the self-phase modulation and the higher-order dispersion will induce perturbations to the soliton dynamics, which lead to the spectral broadening with an access to MIR supercontinuum from a near-infrared pumping pulse \cite{Guo2018NP}. In the experiment, the waveguide with a length of 5 mm allows us to achieve an octave-spanning supercontinuum from 1.5 to 4.2 $\mu$m as illustrated in Fig. \ref{fig2}(e). The average power for the MIR spectral portion is measured about 1.5 mW after a long-pass filter with a cutting wavelength of 2.4 $\mu$m. The pulse duration of the filtered MIR supercontinuum is numerically estimated to be shorter than 1 ps. Details on the waveguide design and supercontinuum simulation can be found in Supplementary Note 2. In contrast to thermal emitters, the supercontinuum source is featured with significantly higher brightness due to the superior spatial coherence, directionality, and beam quality \cite{Zorin2022OE}. 

In parallel, the average power of the YDFL is augmented by two cascaded fiber amplifiers to provide the pump source for the subsequent nonlinear upconversion. The main amplifier is implemented based on photonic crystal gain fibers with a large mode area, which helps to minimize the spectral broadening. As shown in Fig. \ref{fig2}(c), a narrow spectrum with a 3-dB bandwidth of 0.45 nm ($\nu_\text{pump}$ = 4.2 cm$^{-1}$) is obtained at the average power 0.7 W. The corresponding pulse duration is measured to be 9.3 ps as shown in Fig. \ref{fig2}(d). The narrow spectrum of the pump source is essential to obtain a high spectral resolution for the upconversion spectrometer. Additionally, the pulse duration of the pump is much larger than that for the MIR signal, which ensures the signal photons are enwrapped within the center of pump profile to optimize the total conversion efficiency \cite{Barh2019AOP, Huang2021PR}. It is the spectro-temporal engineering for the involved pulses that allows us to realize a nonlinear spectral mapping with a high fidelity and a high efficiency. 

\begin{figure}[b!]
\centering
\includegraphics[width=0.85\columnwidth]{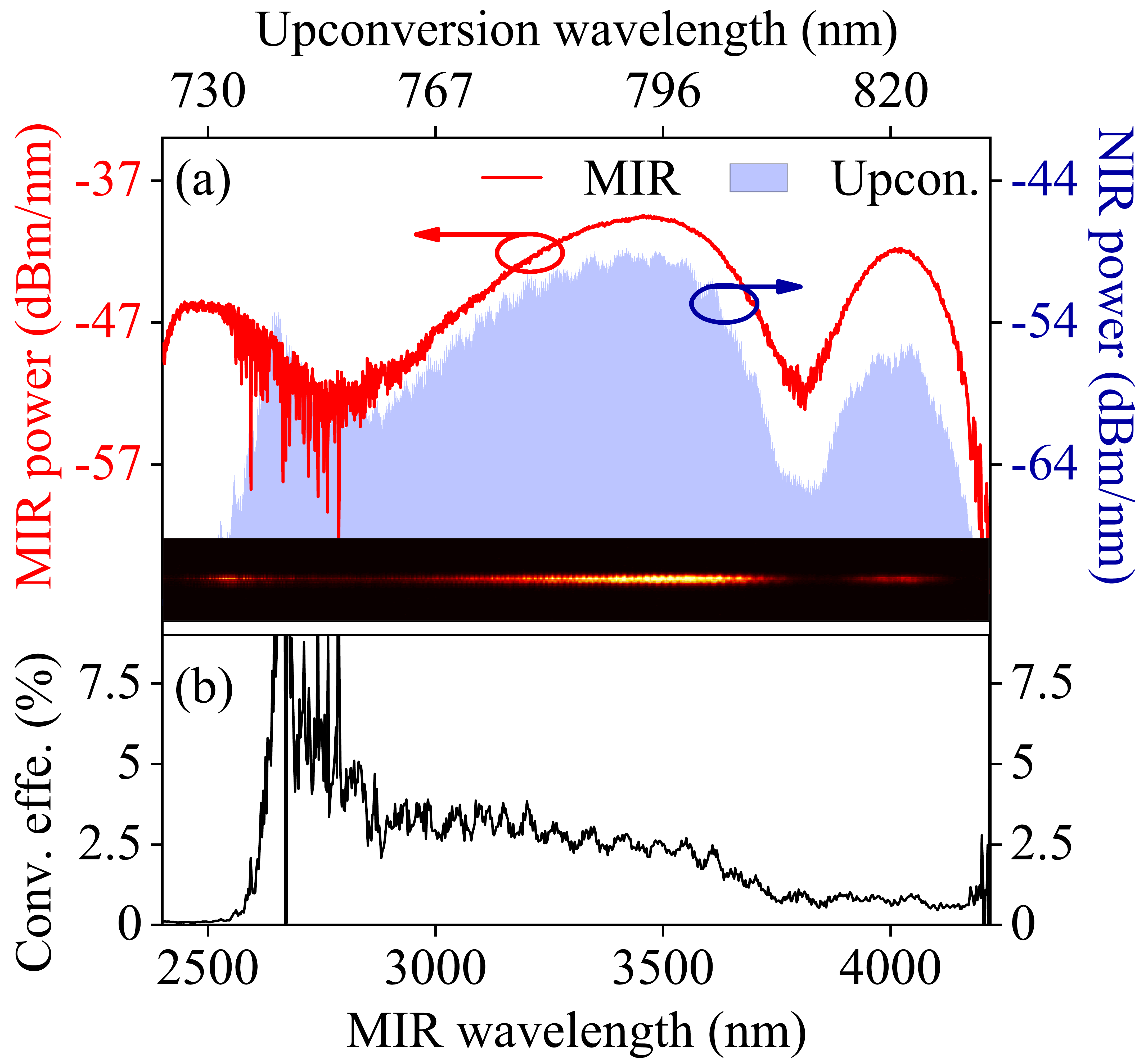}
\caption{Broadband nonlinear frequency upconversion. (a) Measured optical spectrum (red line) for the upconverted signal, which is compared by the input MIR spectral profile (shaded area). The inset presents the upconversion beam dispersed on the silicon camera. (b) Inferred conversion efficiency over the whole band of MIR supercontinuum. The divergence spikes near 2.7 $\mu$m is ascribed to the deficiency power because of water vapor absorptions along the measurement path.}
\label{fig3}
\end{figure}

Then, the MIR signal and the pump sources are spatially combined with a dichroic mirror before being injected into a chirped-poling lithium niobate (CPLN) crystal. The temporal overlap of the mixed beams is required to perform the sum-frequency generation (SFG), which is achieved by carefully tuning the delay line at the path of the pump. The CPLN crystal has a linearly increasing poling period from 19 to 24 $\mu$m along the length of 20 mm. The thickness and width are 1 and 3 mm, respectively. The crystal is operated at room temperature, thus resulting in a simple and robust operation. Moreover, the CPLN allows us to use the collinear configuration to achieve a broadband phase-matching window, thus excluding the spatial chirping effect for the upconverted beam \cite{Wolf2017OE}. As a result, the generated SFG beam can be efficiently coupled into a single-mode fiber, which helps to suppress the ambient scattering noise and pump-induced parametric fluorescence \cite{Huang2021PR} in combination with a series of spectral filters. Finally, the filtered beam is sent into a grating spectrometer that is equipped with a silicon camera for sensitive and fast acquisition of the dispersed photons.

\section{Experimental Results}
\subsection{Broadband MIR upconversion spectroscopy}

Now we turn to characterize the performance of the implemented MIR upconversion spectrometer. The CPLN crystal can support a broadband phase-matching window from 2700 to 5000 nm (3700 to 2000 cm$^{-1}$) due to the self-adapted reciprocal vectors along the light propagating axis \cite{Mrejen2020LPR, Chen2014LSA}, which is significantly wider than the acceptance bandwidth based on periodically-poled lithium niobate (PPLN) crystals \cite{Dam2012NP, Huang2021PR}. Figure \ref{fig3}(a) gives the measured spectrum for the SFG light, along with the recorded camera image by an electron-multiplying charge-coupled device (EMCCDs, Andor iXon Ultra 888), which indicates a simultaneous broadband conversion of the whole MIR supercontinuum. The spectral density $I _\text{SFG}(\lambda_\text{SFG})$ is calibrated with the measured average power of 10 $\mu$W for the SFG beam at the output of the nonlinear crystal. Taking into account the corresponding MIR spectral density $I _\text{MIR}(\lambda_\text{MIR})$, the conversion efficiency defined by the photon-flux ratio is calculated for each spectral element \cite{Neely2012OL, Wolf2017OE} in the non-depletion approximation of the pump field, which reads as $\eta(\lambda_\text{MIR}) = I _\text{SFG}(\lambda_\text{SFG})/I _\text{MIR}(\lambda_\text{MIR}) \times \lambda_\text{SFG}/\lambda_\text{MIR}$, where the SFG wavelength is determined by the energy conservation relationship $\lambda^{-1}_\text{SFG} = \lambda^{-1}_\text{Pump} + \lambda^{-1}_\text{MIR}$. As shown in Fig. \ref{fig3}(b), the conversion efficiency is above 3.2\% for the range between 2.6 to 3.6 $\mu$m. The reduced efficiencies at longer wavelengths may be due to the slightly-diverging pump waists and non-negligible crystal absorption. The spikes around 2.7 $\mu$m is due to the water vapor absorption. It is worth noting that the broadband operation would typically result in a lower conversion efficiency for a given pump intensity. Thanks to the synchronously pulsed pumping, the conversion efficiency has substantially been enhanced by the intensive peak power, which is much higher than previous results based on continuous-wave pumping \cite{Friis2019OL, Jahromi2019OE}. The conversion efficiency can be further increased by augmenting the pump intensity \cite{Mrejen2020LPR}.

\begin{figure}[b!]
\centering
\includegraphics[width=0.75\columnwidth]{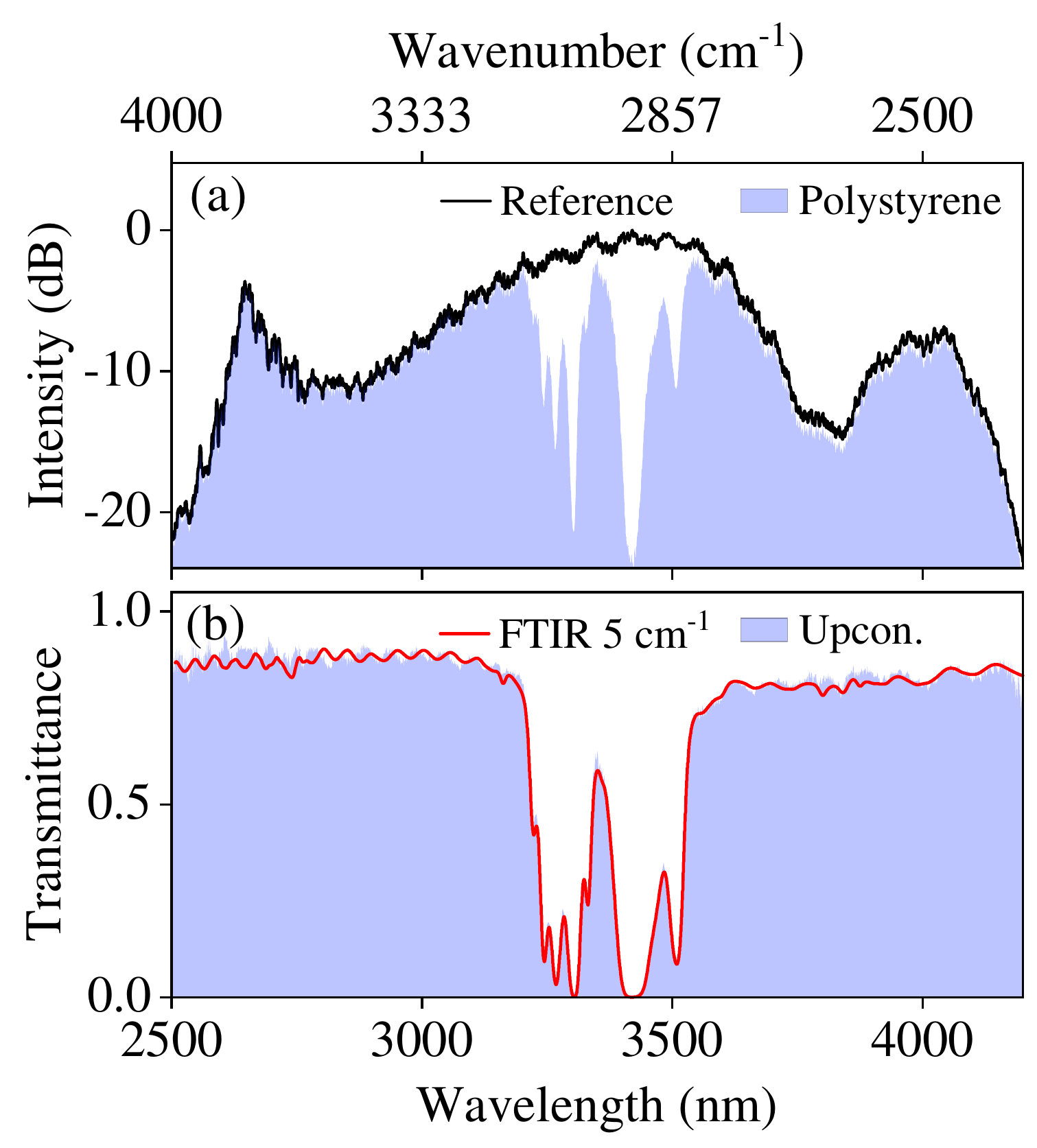}
\caption{Broadband MIR upconversion spectroscopy. (a) Measured MIR spectrum (shaded area) at the presence of a polystyrene film with a thickness of 50 $\mu$m. The reference without the sample is illustrated with the black line. (b) Measured absorption spectrum (shaded area) by the MIR upconversion spectrometer, which is compared to the FTIR spectrum at the resolution setting of 5 cm$^{-1}$.}
\label{fig4}
\end{figure}

In the experiment, the near-infrared spectrometer is calibrated with a mercury argon lamp, which indicates a spectral resolution $\nu_\text{inst}$ of 3.3 cm$^{-1}$. The wavelength values for the measured MIR light are calculated according to the energy conservation law with the knowledge of the pump central wavelength. Figure \ref{fig4}(a) presents the MIR transmissions at the presence and absence of a 50-$\mu$m polystyrene film, which is used to deduce the absorption spectrum of the tested sample. The obtained spectrum agrees well with the one measured by a commercial FTIR spectrometer (Spotlight 400, PerkinElmer) with a spectral resolution of 5 cm$^{-1}$, which verifies the accuracy of our MIR upconversion spectroscopy system. The resolution of the upconversion spectrometer $\nu_\text{MIR-US}$ is determined by the convolution of the optical pump linewidth $\nu_\text{pump}$ and the instrument resolving power $\nu_\text{inst}$, \textit{i.e.}, $\nu_\text{MIR-US} = \sqrt{\nu^2_\text{pump} + \nu^2_\text{inst}}$. The calculated resolution is about 5.3 cm$^{-1}$ closed to the measured value.

\begin{figure}[b!]
\centering
\includegraphics[width=0.85\columnwidth]{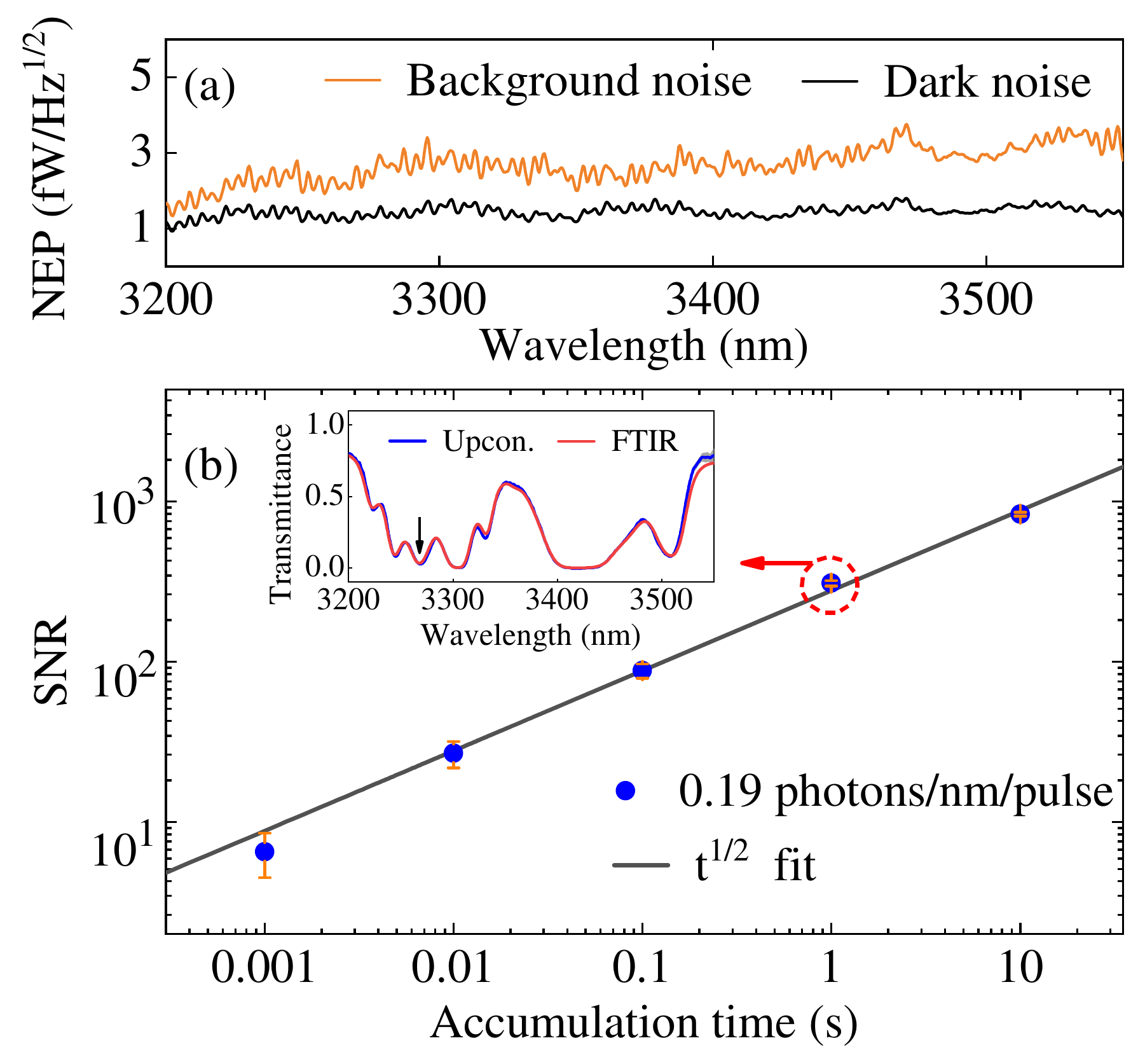}
\caption{Ultra-sensitive MIR spectroscopy at the single-photon level. (a) Noise equivalent power (NEP) at the entrance of the MIR upconversion spectrometer. The background noise is measured at the absence of  MIR illumination, while the dark noise is recorded by further turning off the pump source. (b) Signal-to-noise ratio (SNR) varies as the increase of accumulation time in the case of single-photon-level MIR illumination with 0.19 photons/nm/pulse. The fitted line is given by the square root of the measurement time. Inset presents the recorded MIR absorption spectrum for a 50-$\mu$m polystyrene film under an illumination time of 1 s. The SNR is evaluated at the absorption dip indicated by the arrow.}
\label{fig5}
\end{figure}

\subsection{Single-photon MIR spectroscopy}
Next, we investigate the detection sensitivity of the MIR upconversion spectrometer, which leverages the high-performance silicon-based EMCCD featured with mega-pixel resolution and single-photon responsivity. The detection sensitivity is optimized by operating the EMCCD at a low thermoelectrically-cooled temperature of -80 $^\circ$C and a high on-chip multiplication gain of 1000. Consequently, the dark current of the EMCCD is measured to be about 8$\times$10$^{-4}$  electrons/pixel/second. The power resolving ability for the spectrometer can be evaluated with the noise equivalent power (NEP). In the photon-counting regime, the NEP at the entrance of the upconversion spectrometer is defined as $E(\lambda) \sqrt{2R(\lambda)}/\eta'(\lambda)$, where $E(\lambda)$ is the photon energy at a specific MIR wavelength, $R$ is the measured count rate for the noise, and $\eta'$ is the total detection efficiency \cite{Huang2021PR}. As shown in Fig. \ref{fig5}(a), even at the presence of intense pump excitation, the NEP can be kept at a low level about 3 fW/Hz$^{1/2}$ due to the effective spectral filtering and precise timing gating, which results in about 1000-fold improvements over the $\sim$  pW/Hz$^{1/2}$ level in previous demonstrations \cite{Wolf2017OE}. 

To demonstrate the ultra-sensitive MIR spectroscopy, the MIR illumination is attenuated to the single-photon level with 0.19 photons/nm/pulse. At an accumulation time of 1 s, the recorded photon counts are normalized to give the transmission spectrum for the polystyrene film, showing a SNR of 300. The SNR is defined as the ratio between mean value for the absorption at 3268 nm and the corresponding standard deviation for an assemble of 10 spectral acquisitions. As shown in Fig. \ref{fig5}(b), the SNR increases with the square root of the exposure time. The MIR single-photon spectroscopy would benefit to phototoxicity-free examination of biological specimens and non-destructive characterization of photosensitive chemical materials \cite{Kalashnikov2016NP, Vanselow2020Optica}.

\begin{figure}[b!]
\centering
\includegraphics[width=0.88\columnwidth]{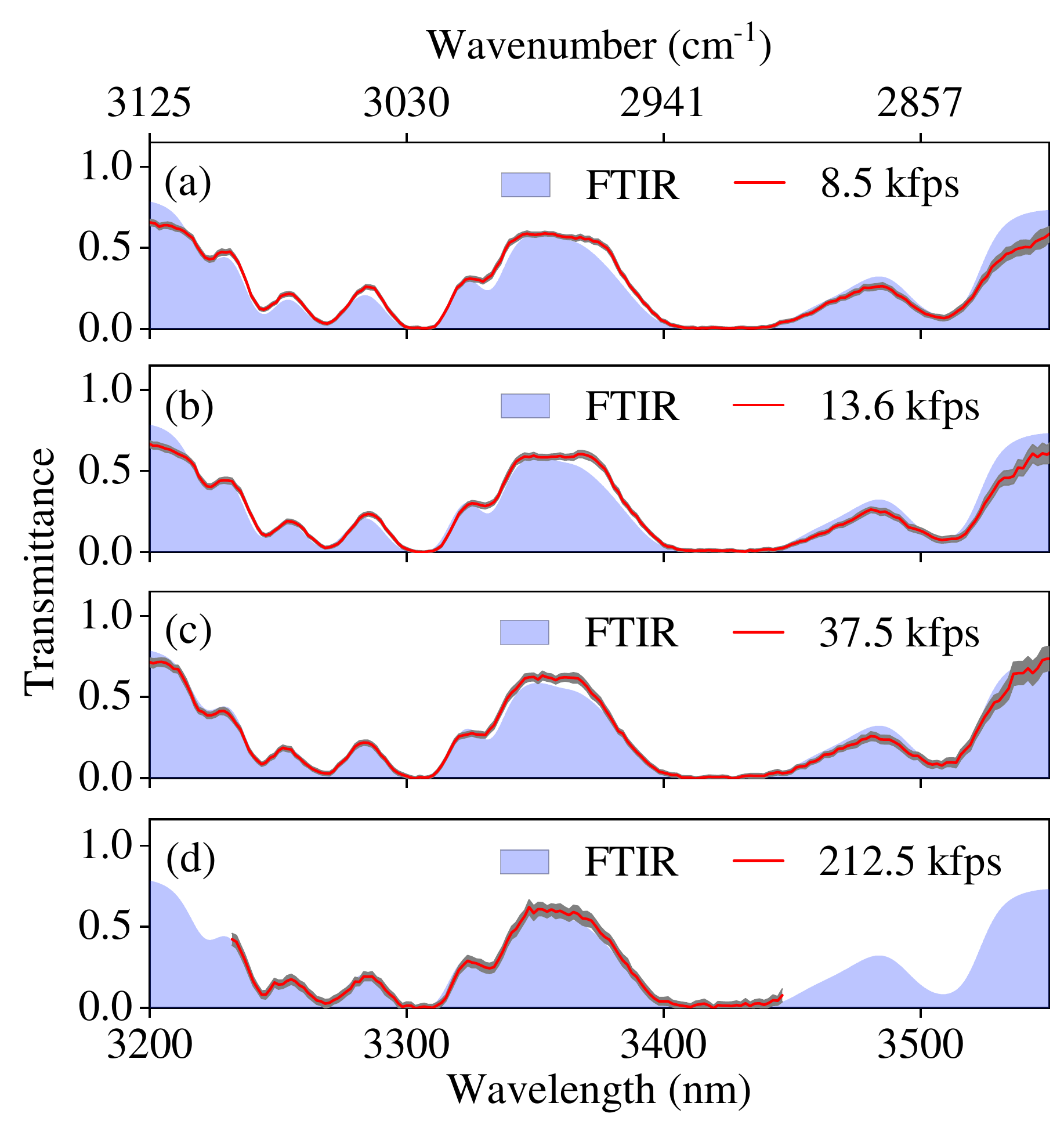}
\caption{High-speed MIR spectroscopy at sub-MHz frame rates. Measured spectra at acquisition rates of 8,500 (a), 13,600 (b), 37,500 (c), and 212,500 (d) frames per second (fps). The shaded area indicates the reference measured by the FTIR with the resolution setting of 5 cm$^{-1}$. The shorter range of the recorded spectrum at a high frame rate is ascribed to the reduced readout pixels of the camera at the high-speed operation.}
\label{fig6}
\end{figure}

\subsection{High-speed MIR spectroscopy}
In the following, we proceed to investigate the high-speed competence of the MIR upconversion spectrometer. The substantial improvement on the detection sensitivity makes it possible to capture the MIR spectrum in a short exposure time, which can significantly improve the frame rate of the spectral acquisition. To this end, a silicon camera (Photron, Mini AX200) based on the complementary metal-oxide semiconductor (CMOS) is used to conduct a fast characterization on the spectroscopic system. Figure \ref{fig6} presents the captured MIR spectra for the polystyrene film at 8.5 k, 13.6 k, 37.5 k, and 212.5 k frames per second, which demonstrate SNRs of 300, 257, 111, and 89 at the absorption peak around 3268 nm, respectively. Note that the single-shot spectral coverage is reduced at high operation rates due to the smaller number of involved pixels, which is technically limited by the speed of readout electronics for the CMOS camera. The frame rate of the MIR upconversion spectrometer can further be enhanced by using linear array detectors that support a readout rate beyond MHz \cite{Rota2016RT}.

\begin{figure}[t!]
\centering
\includegraphics[width=0.95\columnwidth]{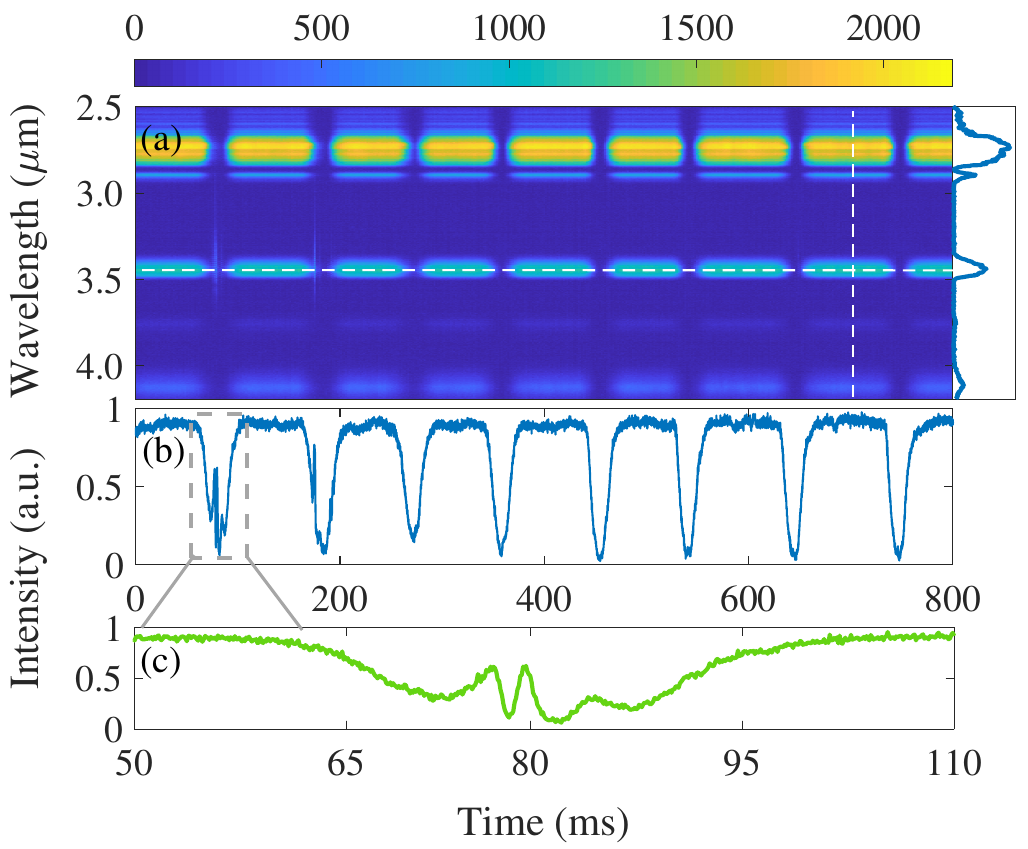}
\caption{Proof-of-principle demonstration of high-speed MIR spectroscopy to monitor the bubble formation process as injecting air into the ethanol. (a) Recorded transmission spectra at the frame rate of 8,500 fps over a measurement time of 800 ms. The section along the vertical dashed line presents a typical spectrum. The transmission at 3490 nm is used to evaluate the air-injection dynamics. The colorbar denotes the measured photon-counting intensity. (b) Transmission trace along the section indicated by the horizontal dashed line in (a). (c) Zoom-in illustration for the early mixing dynamics with more subtle behaviors within 60 ms.}
\label{fig7}
\end{figure}

Finally, a proof-of-principle demonstration is performed to show high-speed condensed-phase MIR spectroscopy. In this case, we monitor the transmission spectrum through a cuvette filled with ethanol, where the air is constantly injected into the liquid to stimulate the mixing dynamics. The real-time spectral acquisition is realized by the MIR upconversion spectrometer at the frame rate of 8.5 kHz. The air-injection dynamics is evaluated by the transmission at 3490 nm as shown in Fig. \ref{fig7}(a). The quasi-periodic perturbation for the bubble formation process is revealed in Fig. \ref{fig7}(b). Intriguingly, the subtle oscillations lasting for several milliseconds are observed as illustrated in Fig. \ref{fig7}(c), which may indicate the initial shock disturbance of the liquid at the beginning of air injection.

\section{Discussions and Conclusion}
The high-speed MIR upconversion spectrometer is made possible by combing the high-brightness MIR illumination, low-noise nonlinear conversion, and high-sensitivity silicon detector. The parallel recording for the dispersed MIR spectrum leads to a spectral acquisition rate up to 212,500 frames per second, which is at least tenfold faster than previous results for mid-IR upconversion spectrometers at similar SNRs \cite{Wolf2017OE, Friis2019OL}. Moreover, the high-speed performance is about one order of magnitude higher than the state-of-the-art scan rates for FTIR-based spectrometers at a comparable spectral resolution \cite{Suss2016RSI, Hashimoto2021LPR, Griffiths1999VS}. To date, the record-high frame rate up to 77 kHz is reported for rapid-scan FTIR spectrometers at the expense of a limited spectral resolution of 9 cm$^{-1}$ \cite{Suss2016RSI}. For a higher spectral resolution of 5.1 cm$^{-1}$, the emerging FTIR variant operates at a rate of 24 kHz based on a phase-controlled delay line \cite{Hashimoto2021LPR}. Therefore, the high-speed and high-resolution performance achieved here have significantly surpassed the reported benchmarks among various MIR spectroscopic techniques (see Supplementary Note 3).

Although the frequency upconversion has long been recognized to implement MIR spectroscopy, yet it remains a long-standing challenge to realize high conversion efficiency and low background noise for a broadband spectral coverage \cite{Gurski1978AO, Rodrigo2021LPR}. So far, the sensitive operation at the single-photon level is only achieved for narrow spectral bandwidths of about tens of nm \cite{Dam2012NP}. And the enlargement of the wavelength acceptance window usually relies on the parameter tuning of phase-matching conditions \cite{Dam2012NP, Junaid2019Optica}. In our experiment, a MIR single-photon upconversion spectrometer is demonstrated with a simultaneous operation band over 1500 nm. The broadband frequency conversion is realized by using a chirped-poling crystal through the adiabatic quasi phase matching \cite{Mrejen2020LPR}. Notably, the nonlinear coincident pumping favors to improve the conversion efficiency due to the high peak power and suppress the background noise via the ultrashort optical gating \cite{Huang2022NC, Huang2021PR}. Consequently, an ultralow noise equivalent power of the spectrometer is manifested, which is about three orders of magnitude improvement over that for previous instantiations \cite{Wolf2017OE}. It is the achieved superior detection sensitivity that enables us to realize high-SNR spectral measurements in the photon-starving regime.

Another distinct feature for the implemented upconversion spectrometer is to leverage the nanophotonic supercontinuum based on a nonlinear Si$_3$N$_4$ waveguide. The chip-based broadband source features a femtosecond pulse duration and a 100-MHz repetition rate, which is in marked contrast to nanosecond supercontinuum sources at MHz-level rates \cite{Jahromi2019OE}. The involved ultrashort duration and ultrahigh rate are in favor of efficiency improvement, noise suppression, and photon accumulation. Moreover, the integrated waveguide provides an efficient and compact platform to approach the broadband MIR coherent light source, comparing to conventional bulky optical parametric oscillators (OPOs) with relatively narrow bands \cite{Hashimoto2021LPR, Junaid2019Optica}. The novel combination with the passively synchronized fiber lasers provides a simple and robust dual-color laser system to facilitate the coincidence-pumping upconversion spectrometer. The synchronous pumping is naturally preferred to obtain higher efficiencies with regard to the pulsed feature of the supercontinuum field, which contrasts with previous implementations based on cavity-enhanced continuous-wave pumping \cite{Jahromi2019OE, Israelsen2019LSA}. In our experiment, the spectro-temporal engineering of the pump pulse allows us to realize the high-efficiency and high-fidelity spectral mapping. The carefully tailoring of the involved optical pulses constitutes the key enabler to achieve the sensitive, fast, and broadband MIR upconversion spectrometer.

Currently, the long wavelength edge of the MIR supercontinuum source is limited to 4200 nm, which is mainly ascribed to the SiO$_2$ cladding of the waveguides. A broader spectral coverage is feasible by using air cladding or sapphire substrates to minimize the propagation loss for optical fields at long infrared wavelengths \cite{Guo2020Optica}. To go beyond the achieved spectral resolution, the pump pulse can be temporally dispersed to implement the chirped pulse upconversion \cite{Zhu2012OE}. In this scenario, only a very narrow frequency component of the pump pulse is involved in the SFG process, which leads to a high-fidelity spectral mapping from the MIR band to the upconverted replica. Additionally, the instrument resolution of the near-infrared spectrometer can be significantly improved to the level below 0.1 cm$^{-1}$ by using enhanced dispersive elements based on virtually imaged phased arrays \cite{Nugent2012OL} or immersion gratings \cite{Iwakuni2019OE}. The extended spectral coverage and enhanced spectral resolution would benefit to facilitate massively parallel sensing of trace gaseous molecules within the functional group region. For instance, the MIR region of 2.5-4.2 $\mu$m hosts the signature of many important greenhouse gases like methane (CH$_4$), nitrous oxide (N$_2$O), and carbon dioxide (CO$_2$).

In summary, we have demonstrated a high-performance MIR upconversion spectrometer based on the supercontinuum source in nanophotonic integrated nonlinear Si$_3$N$_4$ waveguides, which provides an instantaneous operation bandwidth covering 2380 - 3700 cm$^{-1}$. The enhanced detection sensitivity in combination with the high-brightness MIR source allows us to achieve an unprecedented spectral acquisition rate at 212,000 frames per second. The presented high-speed MIR spectrometer is particularly useful to investigate non-repetitive transient phenomena and highly dynamic processes in life, material and chemical sciences, such as high-throughput screening, real-time microfluidic labeling, gaseous combustion measurement, and chemical reaction monitoring.

\begin{backmatter}

\bmsection{Supporting Information} Supporting Information is available from the Wiley Online Library or from the author.

\bmsection{Acknowledgements} This work was supported by National Natural Science Foundation of China (62175064, 62235019, 62035005, 11974234), Shanghai Pilot Program for Basic Research (TQ20220104), Shanghai Municipal Science and Technology Major Project (2019SHZDZX01), Shanghai Science and Technology Development Funds (20QA1403500), and Fundamental Research Funds for the Central Universities.

\bmsection{Conflict of Interest} The authors declare no conflict of interest.

\bmsection{Data Availability Statement} The data that support the findings of this study are available from the corresponding author upon reasonable request.

\bmsection{Keywords} Mid-infrared spectrometer, single-photon spectroscopy, high-speed spectroscopy, frequency upconversion

\end{backmatter}

\end{document}